\documentclass[aps, prx, reprint, longbibliography, nofootinbib,superscriptaddress, floatfix]{revtex4-1}
\usepackage{slashed}
\usepackage{verbatim}
\usepackage[T1]{fontenc}
\usepackage{mathbbol}
\usepackage[dvipsnames]{xcolor}
\usepackage{orcidlink}
\usepackage[ruled,vlined,linesnumbered]{algorithm2e}

\usepackage[normalem]{ulem}
\usepackage[english]{babel}
\usepackage{lipsum}
\usepackage{physics}
\usepackage{dcolumn}
\usepackage{tensor}
\usepackage{comment}
\usepackage{graphicx,color,overpic,mathtools}
\usepackage{tikz}
\usetikzlibrary{shapes.geometric, arrows.meta, positioning, calc}
\usepackage{amsthm,amsmath,amssymb,mathrsfs}
\usepackage{braket,bm,bbm,setspace}
\usepackage{cancel}
\usepackage{float}
\usepackage{xargs}
\definecolor{myred}{RGB}{179, 27, 27}
\usepackage{hyperref}
\hypersetup{
    colorlinks=true,
    linkcolor=myred, 
    citecolor=myred, 
    urlcolor=myred  
 }
\usepackage{siunitx}
\begin{document}

\title{Accurate Black Hole Quasinormal Modes, Regge Poles, and Greybody Factors from Light Ring Ladder Symmetry}

\author{Rajes Ghosh\orcidlink{}}
\email[]{rghosh13@jh.edu}
\affiliation{William H. Miller III Department of Physics \& Astronomy, Johns Hopkins University\\
3400 North Charles Street, Baltimore, MD 21218, USA}
\author{David Pereñiguez\orcidlink{}}
\email[]{dpereni1@jhu.edu}
\affiliation{William H. Miller III Department of Physics \& Astronomy, Johns Hopkins University\\
3400 North Charles Street, Baltimore, MD 21218, USA}
\author{Jaime Redondo-Yuste\orcidlink{0000-0003-3697-0319}}
\email[]{jredondo@princeton.edu}
\affiliation{Leinweber Forum for Theoretical Physics, Princeton University, Princeton NJ 08540, USA}
\affiliation{William H. Miller III Department of Physics \& Astronomy, Johns Hopkins University\\
3400 North Charles Street, Baltimore, MD 21218, USA}
\author{Emanuele Berti\orcidlink{}}
\email[]{berti@jhu.edu}
\affiliation{William H. Miller III Department of Physics \& Astronomy, Johns Hopkins University\\
3400 North Charles Street, Baltimore, MD 21218, USA}

\begin{abstract}
The characteristic oscillations of a black hole can be understood in terms of null rays trapped near its unstable light ring, with this correspondence becoming exact in the high-frequency limit. In this regime, we uncover a hidden ladder symmetry in the perturbation equations, which maps the calculation of quasinormal modes, Regge poles, and greybody factors for a spherically symmetric black hole onto the quantum-mechanical problem of an anharmonic oscillator. This provides a systematic, efficient framework for computing the black hole response, with improved convergence properties compared with existing approximation schemes. As an explicit application, we analytically compute all three quantities for a Schwarzschild black hole through eighteenth order in inverse angular momentum, obtaining highly accurate results. We explore different resummation schemes and find that they can further improve the accuracy of the Regge poles and, particularly, the greybody factors. Our framework can be readily extended to rotating black holes and to modified theories of gravity.
\end{abstract}

\maketitle

\section{Introduction}
The scattering response of black holes (BHs) provides a powerful probe to understand the gravitational wave (GW) emission during the ringdown phase of binary mergers, now observed routinely with rapidly increasing precision~\cite{LIGOScientific:2016aoc, LIGOScientific:2016vlm, LIGOScientific:2020ibl, KAGRA:2021vkt, LIGOScientific:2025pvj, LIGOScientific:2026ctl}. In the time-domain, this response is governed by a series of damped exponentials, called quasinormal modes (QNMs), whose frequencies and damping times encode the properties of the underlying spacetime and provide sensitive tests of general relativity~\cite{Detweiler:1980gk, Kokkotas:1999bd, Dreyer:2003bv, Berti:2005ys, Berti:2009kk, Yunes:2013dva, Berti:2015itd, LIGOScientific:2016lio, Yunes:2016jcc, Berti:2018vdi, Cardoso:2019rvt, Isi:2019aib, Maselli:2019mjd, Maggio:2021ans, Isi:2021iql, LIGOScientific:2021sio, Cotesta:2022pci, Pacilio:2023mvk, Baibhav:2023clw, Franchini:2023eda, Destounis:2023ruj, Maselli:2023khq, Yi:2024elj, Crescimbeni:2024sam, Maenaut:2024oci, Ghosh:2024het, Berti:2025hly, LIGOScientific:2025rid, LIGOScientific:2025wao, Franchini:2025csk, LIGOScientific:2026qni, LIGOScientific:2026oim}. In the frequency-domain, this response is characterized by the BH greybody factor (GBF), which describes the transmission probability of GWs through the effective potential barrier caused by the central BH~\cite{Hawking:1975vcx, Page:1976df, Page:1976ki, Futterman:1988ni, Cvetic:1997ap, Cardoso:2005vb, Harmark:2007jy}. The GBF not only determines the departure of Hawking radiation spectra from an ideal blackbody spectrum, but also provides an accurate description of the frequency-domain waveform. Furthermore, the analytical structure of the scattering problem in the complex angular momentum plane is encoded in Regge poles, which play an important role in wave scattering and gravitational lensing by BHs~\cite{Andersson:1994rk,  Dolan:2008kf, Decanini:2009mu, Folacci:2019vtt}. A unified analytical framework for QNMs, GBFs, and Regge poles is therefore crucial to understand the BH scattering dynamics across the time and frequency domains.

A regime where such an analytical description becomes extremely powerful is the eikonal limit $\ell \gg 1$, where $\ell > |s|$ is the angular momentum quantum number of the perturbing spin-$s$ field~\cite{Press:1971wr, 1972ApJ...172L..95G, Mashhoon:1982im, Blome:1981azp, Ferrari:1984zz, Schutz:1985km, Iyer:1986np, Kokkotas:1988fm, Seidel:1989bp, Konoplya:2003ii, Konoplya:2019hlu, Konoplya:2023moy}. In this limit, the perturbation potential takes a universal form controlled by the unstable light ring (LR) of the spacetime~\cite{Cardoso:2008bp}. This led to the well-known correspondence between the LR geometry and BH spectroscopy, where the BH QNMs are determined, in the eikonal limit, by the orbital frequency and the Lyapunov exponent of the LR~\cite{Blome:1981azp, Mashhoon:1982im, Ferrari:1984zz, Schutz:1985km, Iyer:1986np, Kokkotas:1988fm, Seidel:1989bp, Konoplya:2003ii, Konoplya:2019hlu, Konoplya:2023moy, Cardoso:2008bp}. Several analytical approaches have been developed to exploit this connection, including Wentzel-Kramers-Brillouin (WKB) approximation~\cite{Ferrari:1984zz, Schutz:1985km, Iyer:1986np, Iyer:1986nq, Kokkotas:1988fm, Seidel:1989bp, Kokkotas:1991vz, Konoplya:2003ii, Cardoso:2008bp, Konoplya:2019hlu, Konoplya:2023moy,Tang:2025qaq}, Dolan-Ottewill-type perturbative expansion~\cite{Dolan:2009nk, Yang:2012he, Hatsuda:2021gtn}, and approaches based on Penrose limit~\cite{Fransen:2023eqj, Giataganas:2024hil}. These methods have collectively yielded a rich set of insights and complementary perspectives on the analytic structure of QNMs. 

All of the approaches discussed above agree at leading order, but differ in how post-eikonal corrections are systematically incorporated. These differences lead to distinct convergence properties, reflecting the different ways in which the expansion in $\ell^{-1}$ is organized, and result in substantial variations in both computational efficiency and analytical complexity. For instance, we note that the Penrose limit approach has not been extended beyond linear order, although a sub-leading calculation based on the high-order corrections of~\cite{Blau:2006ar} seems natural. Moreover, the Dolan-Ottewill expansion method has not been applied to compute GBFs beyond linear order. 

Motivated by these gaps, we introduce a complementary approach based on the hidden algebraic symmetry of the near-LR perturbation equation. Our approach may be viewed as a refinement of the Dolan-Ottewill formalism, but with
a key conceptual difference: rather than treating the inverse-angular-momentum expansion solely as an asymptotic expansion of the perturbation equation, we show the appearance of $\mathfrak{sl}(2,\mathbb R)$-symmetry from the underlying action principle. As a result, the leading eikonal problem leads to the construction of harmonic-oscillator type ladder operators. These operators are then systematically used for computing sub-eikonal corrections, which appear as perturbations that break the exact symmetry. 

By invoking quantum mechanical perturbation theory for these sub-leading anharmonic corrections, our method naturally produces an asymptotic expansion in the inverse angular momentum. In this paper we illustrate this structure for the canonical example of a Schwarzschild BH. By carrying the expansion up to 18th order in $(\ell+1/2)^{-1}$, we determine QNMs to considerable accuracy, that improves rapidly for higher angular numbers. Because of the convergence properties of our asymptotic series, resummation techniques (like Padé) do not lead to a significant improvement in the QNM frequencies. On the other hand, by resumming the asymptotic series, we find the Regge poles at low frequencies with an error smaller than $10^{-8}$ for fundamental gravitational perturbation, and sub-percent accurate $\ell=2$ GBF even at high frequencies.

Our algebraic method demonstrates that the problems of determining QNM frequencies, GBFs, and Regge poles, are actually equivalent in the eikonal limit. We show that these are governed by the same equation, only solved for a different variable: the frequency (for QNMs), the complexified angular momentum (for Regge poles), and the analytically continued overtone index (for GBFs). Moreover, this method can be easily extended to parametrized effective potentials, rotating BHs, and phenomenological descriptions capturing deviations from general relativity or the effect of environments~\cite{Tang:2025qaq, Hu:2026cyq, Pezzella:2024tkf}.

An additional appealing feature of this framework is the unifying role played by the underlying ladder symmetry. Remarkably, the same QNM equation that gives rise to a near-horizon ladder structure, previously shown to imply the vanishing of static tidal Love numbers for vacuum BHs in GR~\cite{Hui:2021vcv, Charalambous:2021kcz, BenAchour:2022uqo, Charalambous:2022rre, Sharma:2024hlz, Ghosh:2026vig}, also proves to be a powerful tool for computing sub-eikonal scattering observables. This dual applicability highlights the versatility of the ladder formulation and suggests a deeper structural connection among various observables, emerging as different manifestations of the hidden local symmetries.

The paper is organized as follows. In Section~\ref{sec:eikonal} we demonstrate the emergence of a ladder symmetry in the near-LR expansion of the perturbative equations. In Section~\ref{sec:calculation} we detail our algorithm to compute post-eikonal corrections efficiently up to high orders. Our results and conclusions are summarized in Sections~\ref{sec:results} and~\ref{sec:conclusions}, respectively. Throughout the paper we use geometrical units ($G=c=1$). 

\section{Eikonal Ladder Symmetry}\label{sec:eikonal}

The perturbations of a BH are governed by Schr\"odinger-type equations, when expressed in suitable coordinates. In spherical symmetry, e.g., for a Schwarzschild BH, these equations separate in spin-weighted spherical harmonics, and each $\ell$-multipole of a spin-$s$ perturbation satisfies the Regge-Wheeler equation
\begin{equation}\label{eq:rw}
    \partial^2_{r_\star}\psi + [\omega^2-V_{\ell,s}(r)]\psi = 0 \, , 
\end{equation}
with $\psi$ an appropriate master variable. With $\beta \equiv 1-s^2$, the general potential is 
\begin{equation} \label{V}
    V_{\ell, \beta} = f \Bigl[\frac{\ell(\ell+1)}{r^2}+\frac{2M\beta}{r^3}\Bigr] \, , \quad f = 1-\frac{2M}{r} \, , 
\end{equation}
where $M$ is the BH mass. For instance, the even-parity gravitational perturbations obey this equation with $s=\pm 2$ or $\beta=-3$~\cite{Chandrasekhar:1975nkd,Pereniguez:2026avs}. 

It is convenient to write the potential in terms of $L \equiv (\ell+1/2)$, so that it (and other relevant quantities, as we will see later) takes the form of an asymptotic series in the new variable $L$. Under this redefinition, we have
\begin{equation} \label{pot}
    V_{L,\beta} = \frac{f L^2}{r^2} + \frac{f}{4r^2}\Bigl(\frac{8\beta M}{r}-1\Bigr) \, .
\end{equation}
The strategy now is to investigate the eikonal regime as a systematic expansion in inverse powers of $L$. The physical intuition is that, in this limit, perturbations propagate along null geodesics, and the problem is governed predominantly by the local behavior of the potential about its peak at $r^{\rm peak} = 3M + \mathscr{O}(L^{-1})$, which asymptotically coincides with the LR. This observation suggests several complementary ways of approaching the problem. 

First, since the potential develops a well-defined maximum, one can approximate it locally by a quadratic function (a parabola) and solve the resulting equation near the peak. A global solution is then constructed by matching the local solutions across the turning points. This is essentially the WKB approach~\cite{Ferrari:1984zz, Schutz:1985km, Iyer:1986np, Iyer:1986nq, Kokkotas:1988fm, Seidel:1989bp, Kokkotas:1991vz, Cardoso:2008bp, Tang:2025qaq, Konoplya:2003ii, Konoplya:2019hlu, Konoplya:2023moy}. Secondly, one may exploit the intuition that GWs propagate along null geodesics, and factor out the leading radial Hamilton-Jacobi phase from the wavefunction. This automatically captures the boundary conditions, and reduces the problem to a global differential equation that admits a perturbative solution in inverse powers of $L$. This approach was first pursued by Dolan and Ottewill~\cite{Dolan:2009nk}, and then extended to rotating BHs by Dolan~\cite{Dolan:2010wr}. 

In contrast, our approach here consists of zooming into the spacetime close to the peak, rescaling the radial coordinate in an appropriate manner. In spirit, this resembles taking the Penrose limit of the spacetime close to the LR~\cite{Fransen:2023eqj}, but our construction is performed entirely at the level of the master equation. As we will show, the resulting expansion reproduces exactly the results of Ref.~\cite{Dolan:2009nk}, while also revealing hidden symmetries which can be exploited to speed up the computation. 

Let us start by redefining the tortoise coordinate as 
\begin{equation}
    \delta r_*= r_*-r_{*}^{\rm peak} \equiv C_L\, y \, .
\end{equation}
Here $C_L$ is a multiplicative factor depending on $L$, that for the moment we treat as a free parameter. Plugging this redefinition into Eq.~\eqref{eq:rw}, and keeping only the leading order in $L$, we have 
\begin{equation}
    \frac{d^2\psi}{dy^2} + \frac{1}{2}\, C_L^4\, \Bigl|\frac{d^2 V}{dr_*^2}\Bigr|_{\rm peak}\, y^2\, \psi = -C_L^2\, (\omega^2-V^{\rm eik})\, \psi \, .
\end{equation}
The curvature of the potential at the peak is written simply in terms of the LR frequency $\Omega_c$ and Lyapunov exponent $\lambda_c$ as $|{d^2 V}/{dr_*^2}|_{\rm peak}=2L^2 \Omega_c^2\lambda_c^2$. The square of the LR frequency is defined as the height of the potential maximum, up to a factor of the angular momentum, i.e., $\Omega_c^2= V^{\rm eik}/L^2$. For a Schwarzschild BH, it coincides with the Lyapunov exponent, $\lambda_c=\Omega_c=1/(3\sqrt{3}M)$. We notice that choosing
\begin{equation}
    C_L^2 = \frac{\mathrm{i}}{L\, \Omega_c\, \lambda_c}
\end{equation}
reduces the leading-order near-LR Regge-Wheeler equation to
\begin{equation}
    H_{\textrm{eik}}\,\psi = \mathcal{E}^{\rm eik}_{\omega,L}\,\psi \, , \quad \mathcal{E}^{\rm eik}_{\omega,L} = (\omega^2-L^2\,\Omega_c^2)\, C_L^2 \, ,
\end{equation}
with $H_{\textrm{eik}}=(-\partial^2_y + y^2)$, resembling the structure of a harmonic oscillator. Hence, the different energy levels will naturally correspond to quantized values of $\mathcal{E}_{\omega,L}$, or equivalently, to a quantization of the frequencies. These are the different \emph{overtones}, labeled by an index $n=0,\,\dots,\,\infty$ for each $L$ sector. Therefore we can label these quantized values of the energy as $\mathcal{E}_{n,L} \equiv \mathcal{E}_{\omega_{Ln},L}$.

The above differential equation supports ladder operators $D^{\pm}$ that raise and lower the overtone index $n$. Following the prescription of Ref.~\cite{Ghosh:2026vig}, these operators and their energy eigenvalues can be calculated as
\begin{equation} \label{ladder}
    \begin{split}
        D^\pm = \mp \partial_y+y,\quad \mathcal{E}_{n L}=2 n+1 \, .
    \end{split}
\end{equation}
In fact, one obtains $\mathcal{E}_{nL}=\pm 2 n+a_L$. Requiring that the modes are stable, i.e., that they correspond to QNMs, selects the positive branch, which leads to negative imaginary parts for the QNM frequencies. Moreover, the constant $a_L=1$ is fixed by demanding the existence of a ground state satisfying $D^{-}|0\rangle=0$. Defining the auxiliary variable $N \equiv (n+1/2)$, this means that the energy eigenvalues are independent of $L$, 
\begin{equation}
    \mathcal{E}_N = 2N \implies \omega^{\rm eik}_{LN} = L\, \Omega_c-\mathrm{i}\, N\, \lambda_c \, .
\end{equation}
Note also that the ladder operators are Hermitian conjugates of each other, i.e., $(D^\pm)^\dagger=D^\mp$. For future reference, we also highlight some key identities:
\begin{equation} \label{id}
    \begin{split}
        &H_{\text{eik}}=D^+ D^-+1, \quad \left[D^+,D^-\right]=-2,\\
        &D^+ |N \rangle = \sqrt{2N+1}\, |N+1\rangle,\\
        &D^-|N \rangle = \sqrt{2N-1}\, |N-1\rangle,
    \end{split}
\end{equation}
where the last two relations can be verified using the fact that the state $|N\rangle$ indeed satisfies $H_{\text{eik}}\, |N\rangle = 2N\, |N\rangle$. From these identities and $y = \left(D^++D^-\right)/2$, it is also easy to check that $\langle y^{\text{odd}} \rangle_N \equiv \langle N | y^{\text{odd}} | N \rangle = 0$ for any odd powers of $y$, and $\langle y^2 \rangle_N \equiv \langle N | y^2 | N \rangle = N$. Moreover, note that the state $|N\rangle$ is properly normalized as $|| D^- | N \rangle||^2=(2N-1)$ by using the last relation in Eq.~\eqref{id}, which is consistent with $\langle N | D^+D^- | N \rangle = \langle N | H_{\text{eik}}-1 | N \rangle$. 

Interestingly, the origin of this eikonal ladder structure can be seen already at the level of the action. Consider a minimally coupled, massless scalar field $\Phi$, whose action is given by 
\begin{equation} \label{action1}
    S[\Phi]=\frac{1}{2}\int d^4x\, \sqrt{-g}\, \Phi\, \Box_{\text{Sch}}\, \Phi.
\end{equation}
Here, we have rewritten the more recognizable form $g^{\mu \nu}_{\text{Sch}} \nabla_\mu \Phi\, \nabla_\nu \Phi$ in terms of $\Phi\, \Box_{\text{Sch}}\, \Phi$ using integration by parts and neglecting the surface terms. We can decompose the field in spherical harmonic modes $\Phi= \sum \psi_{\ell m}Y_{\ell m}$, and reduce Eq.~\eqref{action1} to a $(1+1)$-dimensional action 
\begin{equation} \label{action2}
    S[\Phi]=\frac{1}{2}\, \sum_{\ell m}\, \int dt dr_*\,\psi_{\ell m} \left[-\partial_t^2+\partial_{r_*}^2-V(r) \right]\psi_{\ell m}\, ,
\end{equation}
where we have used the definition of the tortoise coordinate $dr=f(r)\, dr_*$ and the potential $V(r)$ is given by Eq.~\eqref{pot}, or equivalently by Eq.~\eqref{V}. 

In the eikonal limit, the action is dominated by the contributions with higher frequency, $L \gg 1$. Therefore we can safely ignore the lower-frequency terms, and study (to leading order) the eikonal action
\begin{equation} \label{Seik}
    S_{\text{eik}}[\psi] \approx \frac{1}{2} \sum_{L\gg1, m} \int \frac{dtdy}{C_L}\left[\psi_{Lm} \mathcal{T} \psi_{Lm}-\psi_{Lm} H_{\text{eik}} \psi_{Lm}\right],
\end{equation}
where we recall that $L=\ell+1/2$, 
$\mathcal{T} \equiv C_L^2(-\partial_t^2-V_{\text{eik}})$ is a time-domain operator, and the spatial operator is $H_{\textrm{eik}}=(-\partial^2_y + y^2)$. 

The above equation represents the leading-order action governing fluctuations localized near the LR. Its temporal sector determines the eigenvalues $\omega$, whereas the ladder structure is encoded in the transverse operator $H_{\text{eik}}$. In terms of the momentum operator $P=-\mathrm{i}\, \partial_y$ conjugate to $y$ so that $[y,P]=\mathrm{i}$, the eikonal Hamiltonian becomes $H_{\text{eik}}=P^2+y^2$. Now, let us introduce three quadratic operators as
\begin{equation} \label{ops}
    K_+=\frac{y^2}{2},\, \, K_{-}=\frac{P^2}{2},\, \, K_0=-\frac{\{y,P\}}{4}.
\end{equation}
They obey the standard $\mathfrak{sl}(2,\mathbb R)$ algebra~\cite{Hadar:2022xag}: $[K_0,K_+]=\mathrm{i}\, K_+$, $[K_0,K_-]=-\mathrm{i}\, K_-$, and $[K_+,K_-]=-2\,\mathrm{i}\, K_0$. Furthermore, $H_{\text{eik}}$ can be written as a linear combination of the ``generators'' $K_\pm$ as $H_{\text{eik}}=2(K_++K_-)$. Therefore, the emergent dynamical $\mathfrak{sl}(2,\mathbb R)$ algebra is equivalent to the quadratic combinations of the ladder operators defined in Eq.~\eqref{ladder}, i.e., $\mathfrak{sl}(2,\mathbb R) \sim \{(D^\pm)^2, D^+D^-\}$, which naturally labels the overtone spectrum (as shown above). Although the above discussion refers to a massless scalar field, similar conclusions naturally follow for higher-spin fields in the Schwarzschild background.

\section{Systematic Calculation of Post-Eikonal Corrections}\label{sec:calculation}

The above discussion illustrates the emergence of a ladder symmetry structure in the leading-order, eikonal equation zoomed-in near the LR. We will now expand the equation to an arbitrarily high order in $1/L$ and show that the problem reduces to a deformed harmonic oscillator. As a result, tools from perturbation theory in quantum mechanics can be readily applied to this problem, finding analytic solutions for the post-eikonal corrections to the QNM frequencies, Regge poles, and GBFs. 

First, it is useful to note that the peak of the potential $r^{\rm peak}$ admits a closed form expression
\begin{equation}
    \begin{split}
        r^{\rm peak}= &\frac{3M}{2\left(L^2-\frac{1}{4}\right)}\left[\left(L^2-\frac{1}{4}\right)-\beta+\left\{\left(L^2-\frac{1}{4}\right)^2 \right. \right.\\
        &\kern5em \left. \left.+\frac{14\beta}{9}\left(L^2-\frac{1}{4}\right)+ \beta^2\right\}^{1/2} \right] \, .
    \end{split}
\end{equation}
As $L \to \infty$, then $r^{\rm peak} \to 3M$ as expected. Writing Eq.~\eqref{eq:rw} in terms of the rescaled coordinate $y = (r_* - r_*^{\rm peak})/C_L$, and keeping $\omega$ as a free parameter, we obtain 
\begin{equation} \label{Hk}
    \Biggl(H_{\rm eik} + \sum_{k\in\mathbb{N}/2}\frac{H_k(y)}{L^k}\Biggr)\psi = \mathcal{E}(\omega,L)\, \psi \, ,
\end{equation}
where we recall that $H_{\rm eik} = (-\partial^2_y + y^2)$, and the sub-leading corrections appear as anharmonic perturbations to the harmonic oscillator. For reference, we list here a few sub-leading corrections:
\begin{equation}
    \begin{aligned}
        &H_{1/2} = (1+\mathrm{i})\sqrt{\frac{2}{27}}y^3, \,\, H_1 = - \frac{2 \mathrm{i} y^4}{3} \, , \\
        &H_{3/2} = \frac{(\mathrm{i}-1)}{\sqrt{6}}y^5,\, \, H_2 =-\frac{45-160\beta+48y^4}{180}y^2 \, ,  
    \end{aligned}
\end{equation}
and we capture all the \emph{constant} ($y$-independent) coefficients in $\mathcal{E}(\omega,L)$, obtaining to $O(L^{-2})$
\begin{equation}
    \mathcal{E}(\omega,L) = \mathcal{E}_{\omega,L}^{\rm eik}  + \frac{\mathrm{i}}{4L}\left(1-\frac{8\beta}{3}\right) + \mathscr{O}(L^{-2}) \, .
\end{equation}
This strategy can be easily extended to higher powers in $1/L$ as well. Now, requiring the eigenvalues to be quantized yields, in the usual manner, the corrections to the energy eigenvalue due to the anharmonic terms
\begin{equation}
    \mathcal{E}_N = 2N + \sum_{k} \frac{\epsilon_k(N)}{L^k} \, .
\end{equation}
In fact, requiring that the eigenfrequencies of the perturbed harmonic oscillator coincide with the constant term of the equation $\mathcal{E}(\omega,L) = \mathcal{E}_N$ establishes an algebraic equation for the QNM frequencies, or determines the Regge poles by complexifying $L$. Moreover, as we will show later, this also allows us to compute the GBFs. In the following we detail our implementation. 

\subsection{QNM Frequencies}
In order to determine the QNM frequencies, we expand the frequency $\omega$ in powers of $1/L$, keeping $\{N,\beta\}$ fixed:
\begin{equation} \label{QNMser}
    \omega= \omega^{\rm eik}_{LN} + \sum_{k=1}^\infty \frac{\omega^{(k)}_N}{L^k} \, , 
\end{equation}
where $\omega^{(k)}$ are the coefficients to determine. Notice that we do not include any half-integer powers. Indeed, the half-integer anharmonic corrections are odd, scaling as 
\begin{equation}
    H_{(2k-1)/2} \propto y^{2k+1} \, .
\end{equation}
Therefore, the diagonal elements $\langle N|H_{(2k-1)/2)}|N\rangle$ trivially vanish for all $N$, and so do the products of the expectation values of an \emph{odd} number of such terms. This, however, does not mean that these anharmonic corrections do not contribute at all to the QNM frequencies: they do, but they enter only at integer orders in $1/L$. 

The strategy of the calculation is rather straightforward. Let us consider the first correction, $\omega^{(1)}_N$. In order to compute it, we need to know $\epsilon_{(1/2)}$ and $\epsilon_1$. Following our argument above, $\epsilon_{1/2}\sim \langle N|H_{1/2}|N\rangle$ vanishes identically. The next order is given by 
\begin{equation}
    \begin{aligned}
        &\epsilon_1(N) = \langle N|H_1|N\rangle + \sum_{N'\neq N}\frac{|\langle N|H_{1/2}|N'\rangle|^2}{2(N-N')} \\
        &\kern2.5em =-\frac{\mathrm{i}}{216}(61+276N^2) \, .
    \end{aligned}
\end{equation}
Solving now for $\mathcal{E}(\omega_n,L) = \mathcal{E}_N$ leads to 
\begin{equation}
    \sqrt{27}M \omega^{(1)}_N = \frac{\beta}{3}-\frac{5N^2}{36} - \frac{115}{432} \, , 
\end{equation}
in agreement with Ref.~\cite{Dolan:2009nk}. 

This expansion can be carried out to very high order by including contributions of suitable $H_k$ terms, each entering at a different perturbative order. There are two such contributions at first order, but the number of contributions grows rapidly. In fact, the total number of terms contributing to $\omega^{(k)}_N$ is equal to the number of ways $k$ can be written as a sum of positive half-integers, that is, the partition of $2k$, which we denote by $p(2k)$. The Ramanujan-Hardy formula now implies that $p(2k)\sim (8k\sqrt{3})^{-1}\exp[\pi \sqrt{4k/3}]$~\cite{andrews1976theory}, i.e., the number of terms grows almost exponentially with the perturbative order. Thus, it is important to devise efficient algorithms to compute them. We have implemented two such methods and cross-checked them, finding excellent agreement. 

\subsubsection{Bender-Wu Implementation}

The first method makes use of the \texttt{BenderWu} (BW) package developed in Ref.~\cite{Sulejmanpasic:2016fwr}, which builds upon the seminal work of Ref.~\cite{Bender:1973rz}. The BW algorithm generates high-order energy and wavefunction corrections due to anharmonic perturbations around a locally harmonic potential through an elegant recursive reformulation of standard perturbation theory. Its implementation, however, requires the perturbing potential to be expressed in a particular power-series form. 

The BW package is designed for Hamiltonians describing anharmonic perturbations of a harmonic oscillator with the general form
\begin{equation} \label{HBW}
    H_{BW}(x) = -\frac{1}{2} \partial_x^2 + \frac{1}{2}x^2 + \sum_{k \geq 3} v_k(g)\, x^k\, , 
\end{equation}
where $x$ is a dimensionless coordinate quantifying the distance from the extremum ($x=0$) of the potential, and $v_k(g)$ are the coefficients of the anharmonic terms, that depend on the ``coupling'' $g$ as some power law. 

By construction, the coefficients of the constant, linear, and quadratic terms are taken to be zero, i.e., $v_0(g)= v_1(g)=v_2(g)=0$. The condition $v_1(g)=0$ simply reflects the choice of expansion point at the extremum of the potential. The absence of $\{v_0(g),v_2(g)\}$ are equally natural, as the presence of $v_0(g)$ will merely shift all energy levels uniformly, whereas a quadratic correction $v_2(g)\, x^2$ can be absorbed into a redefinition of the harmonic oscillator frequency. Consequently, the BW formalism considers a canonical harmonic oscillator perturbed solely by higher-order anharmonic interactions.

However, for our case of sub-eikonal QNMs, the Hamiltonian given by Eq.~\eqref{Hk} contains perturbative contributions proportional to $y^2$. Therefore, before the BW machinery can be applied, these terms must be eliminated. This is accomplished by introducing a coordinate transformation $y \to y(x) = x\, \tilde{y}(g)$ that depends on the coupling $g \equiv 1/L$ as
\begin{equation} \label{ty}
    \tilde{y}(g) = 1+ \sum_{k \in 2\mathbb{N}} x_k\, g^k,
\end{equation}
where the $x_k$'s are chosen so that the coefficient of $x^2$ in the transformed Hamiltonian becomes $1/2$. Solving for the $x_k$'s order-by-order in $g$, we can reduce the Hamiltonian to the canonical form given by Eq.~\eqref{HBW}. We can now implement this Hamiltonian in the BW package and compute the energy shifts, which will, in turn, provide the $\omega_k^{(n)}$'s. We have systematically performed this procedure in \texttt{Mathematica} and computed terms up to $\omega_4^{(n)}$, finding that all of them are in agreement with Ref.~\cite{Dolan:2009nk}.

\subsubsection{Recursive Ladder Implementation}

The BW algorithm provides an efficient route for computing high-order perturbative corrections, but it is instructive to develop an alternative implementation that works directly with the structure of the eikonal expansion given by Eq.~\eqref{Hk}. Besides serving as an independent check of the BW results, this construction makes the underlying perturbative organization transparent and avoids the need for intermediate coordinate transformations.

Here we simply make use of the fact that the potential is a power law in $y$, and that the action of $y$ on particle number eigenstates is known. Let us denote 
\begin{equation}
    \delta H = \sum_{k=1} \delta H_k\, L^{-k/2} = \sum_{k=1}\sum_{j=0} h_{k,j}\, y^j\, L^{-k/2} 
\end{equation}
for some coefficients $h_{k,j}$. We wish to find the perturbed energy eigenvalues
\begin{equation}
    (H_{\rm eik}+\delta H)\ket{\psi} = E\ket{\psi} \, .
\end{equation}
The eigenstates are also perturbed as 
\begin{equation}
    \ket{\psi} = \ket{N} + \sum_{k=1} \ket{\psi^{(k)}} \, , 
\end{equation}
where normalization requires that $\braket{N|\psi^{(k)}}=0$. We solve order-by-order for (i) the corrections to the energy $\epsilon^{(k)}$, and (ii) the corrections to the eigenstate $\psi^{(k)}$. The $k$-th order energy eigenvalue shift is simply 
\begin{equation}
    \epsilon^{(k)}=\sum_{j=1}^k\braket{N |\delta H_j|\psi^{(k-j)}}  \, .
\end{equation}
Next, we can find the eigenstate correction at order $k$ via 
\begin{equation}
    \ket{\psi^{(k)}} = - \sum_{j\neq 0}\sum_{l=1}^k \frac{\braket{N+j|\delta H_k|\psi^{(k-l)}}}{2j}\ket{N+j} \, .
\end{equation}
Above we are using that $E_0(N)-E_0(N+j)=-2j$. This approach admits a straightforward algorithmic implementation in \texttt{Mathematica}, which we have pushed up to $\mathscr{O}(L^{-18})$. Here, we only quote the result for the fundamental gravitational QNM with $\{N=1/2,\, \beta=-3\}$:
\begin{widetext}
\begin{equation*}
    \begin{split}
    &\sqrt{27}\, M\, \omega = L - \frac{\mathrm{i}}{2} - \frac{281}{216\, L} + \frac{1591\, \mathrm{i}}{7776\, L^{2}} - \frac{710185}{1259712\, L^{3}} + \frac{92347783\, \mathrm{i}}{362797056\, L^{4}} - \frac{7827932509}{39182082048\, L^{5}}- \frac{481407154423\, \mathrm{i}}{8463329722368\, L^{6}}\\
    &+ \frac{133552310100467}{914039610015744\, L^{7}} - \frac{307863728279673163\, \mathrm{i}}{263243407684534272\, L^{8}} + \frac{11457725853104647541}{255872592269367312384\, L^{9}} - \frac{64399920131779549902077\, \mathrm{i}}{18422826643394446491648\, L^{10}}\\
    &- \frac{2990345419943957075918059}{1989665277486600221097984\, L^{11}} - \frac{5640143930052247233866116621\, \mathrm{i}}{859535399874211295514329088\, L^{12}} - \frac{408538462479995155399829897833}{92829823186414819915547541504\, L^{13}} \\
    &- \frac{69656132994811319310158940504613\, \mathrm{i}}{6683747269421867033919422988288\, L^{14}} - \frac{15789486889600735523221656517052981}{2165534115292684918989893048205312\, L^{15}}\\
    &- \frac{18976930821707554997846963339150903243\, \mathrm{i}}{1247347650408586513338178395766259712\, L^{16}} - \frac{306490327764337574850494609370856264651}{134713546244127343440523266742756048896\, L^{17}}\\
    &- \frac{1622351165560128953704767763386569023267201\, \mathrm{i}}{87294377966194518549459076849305919684608\, L^{18}} + \mathscr{O}\!\left(L^{-19}\right)\,.
    \end{split}
\end{equation*}
\end{widetext}
In general, we notice a few algebraic properties of the asymptotic QNM  series: $\omega(-N,\, L)=\omega^*(N,\, L)$, $\omega(N,\, -L)=-\omega^*(N,\, L)$, and hence the series is odd under a simultaneous inversion $\{N,L\} \to \{-N,-L\}$. These relations provide nontrivial constraints on the allowed structure of the series coefficients.

\subsection{Regge Poles}

Regge poles are defined as the poles of the scattering matrix in the complexified angular-momentum plane. To obtain them within our framework, we analytically continue the angular-momentum parameter from its physical integer values to the complex plane, $L \to \lambda_\omega^{(n)} \in \mathbb{C}$, while keeping the frequency $\omega$ real. Here, $n$ labels the different Regge trajectories. In other words, we solve the truncated equation
\begin{equation}
    \mathcal{E}(\omega,\lambda_\omega^{(n)}) = \mathcal{E}_n^{\rm eik}(L \to \lambda_\omega^{(n)}) + \sum_{j=1}^k \frac{\epsilon_j(N)}{[\lambda_\omega^{(n)}]^j} \,
\end{equation}
for $\lambda_\omega^{(n)}$. This yields a $k$-th order accurate approximation to the $N=(n+1/2)$-th Regge pole, at frequency $\omega$. The lower orders are given simply by 
\begin{equation}
    \begin{split}
        &\lambda_\omega^{(n)} = \varpi+\mathrm{i}\, N+\frac{60N^2-114\beta+115}{432\,\varpi} \\
        &\kern3em -\mathrm{i}\, N \frac{(1220 N^2- 6912 \beta + 5555)}{15552\, \varpi^2}+\cdots\,,
    \end{split}
\end{equation}
where we have used the shorthand $\varpi \equiv \sqrt{27} M\omega$. The above expression is in agreement with Ref.~\cite{Dolan:2009nk}. Notice how the Regge poles $\lambda_\omega^{(n)}$ can equivalently be obtained by inverting the asymptotic series expansion in Eq.~\eqref{QNMser} for the QNM frequencies and solving for complexified-$L$ in terms of $\omega$. 

We have computed the expression of the Regge poles, exact in $\{N,\beta\}$, up to order $\mathscr{O}(L^{-18})$ and we find the following algebraic identities: $\lambda_{-\omega}^{(N)}=-\lambda_\omega^{*(N)}$, $\lambda_\omega^{(-N)}=\lambda_\omega^{*(N)}$, hence the series is odd under a simultaneous inversion $\{N,\omega\} \to \{-N,-\omega\}$. Here, for brevity, we only quote the result for the fundamental gravitational Regge pole with $\{N=1/2,\, \beta=-3\}$:

\begin{widetext}
    \begin{equation*}
        \begin{split}
            &\lambda_{\omega}^{(n=0)} = \varpi + \frac{\mathrm{i}}{2} + \frac{281}{216\, \varpi} - \frac{6649\, \mathrm{i}}{7776\, \varpi^{2}} - \frac{1044601}{629856\, \varpi^{3}} + \frac{926224193\, \mathrm{i}}{362797056\, \varpi^{4}} + \frac{184851431845}{39182082048\, \varpi^{5}} - \frac{71361067332161\,\mathrm{i}}{8463329722368\, \varpi^{6}}\\
            &- \frac{14390928366797903}{914039610015744\, \varpi^{7}} + \frac{7717840397981223883\,\mathrm{i}}{263243407684534272\, \varpi^{8}} + \frac{14117571610293670714747}{255872592269367312384\, \varpi^{9}} - \frac{1907938000811368419778219\,\mathrm{i}}{18422826643394446491648\, \varpi^{10}} \\
            &- \frac{385765536623517642670935281}{1989665277486600221097984\, \varpi^{11}} + \frac{308973753349472755029744697069\,\mathrm{i}}{859535399874211295514329088\, \varpi^{12}} + \frac{60901535701933258231377209547649}{92829823186414819915547541504\, \varpi^{13}}\\
            &- \frac{7783297193742044311750741091591843\,\mathrm{i}}{6683747269421867033919422988288\, \varpi^{14}} - \frac{4260829605643457340038346820111576711}{2165534115292684918989893048205312\, \varpi^{15}}\\
            &+ \frac{3757519891398658340844939128329326463691\,\mathrm{i}}{1247347650408586513338178395766259712\,\varpi^{16}} + \frac{483296066678553440178719144747613201690283}{134713546244127343440523266742756048896\,\varpi^{17}}\\
            &- \frac{49350957965236608416609703502382780677890887\,\mathrm{i}}{87294377966194518549459076849305919684608\, \varpi^{18}} + \mathscr{O}\!\left(\varpi^{-19}\right)\,.
        \end{split}    
    \end{equation*}
\end{widetext}
Note that the imaginary part of the Regge pole encodes the attenuation of the corresponding surface-wave contribution as the wave propagates around the BH, while its real part determines the associated angular momentum. The resulting trajectories $\lambda_\omega^{(n)}$ therefore provide a complementary characterization of the same barrier-scattering problem encoded in the QNM spectrum.

\subsection{Greybody Factors}

Finally, we demonstrate that the same ladder construction can be used to determine systematically the BH reflectivity and GBFs. This is a novelty of our work: previous approximations to the greybody factors relied only on WKB approximations~\cite{Iyer:1986np, Iyer:1986nq}, and on resummations thereof in terms of QNM frequencies~\cite{Konoplya:2019hlu, Konoplya:2023moy}. 

Consider the scattering of monochromatic waves with frequency $\omega$ and angular number $\ell$ on the BH. Asymptotically, the master variable (equivalently, the GW strain, up to a numerical factor) behaves as 
\begin{equation}
    \psi = A_{\rm in}(\omega,\ell)e^{-i\omega r_*}+A_{\rm out}(\omega,\ell)e^{i\omega r_*} \, ,
\end{equation}
where $A_{\rm in}$ and $A_{\rm out}$ denote the amplitudes of the incoming and reflected waves, respectively. The BH reflectivity is, therefore, defined to be the quotient between the outgoing and ingoing amplitudes
\begin{equation}
    \mathcal{R}_\ell(\omega) = \Bigl|\frac{A_{\rm out}(\omega,\ell)}{A_{\rm in}(\omega,\ell)}\Bigr|^2 \, .
\end{equation}
For a BH horizon, imposing purely ingoing boundary conditions, flux conservation gives the corresponding greybody factor $\Gamma_\ell(\omega)=1-\mathcal{R}_\ell(\omega)$. Thus, $\Gamma_\ell$ measures the probability for the incident wave to penetrate the potential barrier and be absorbed by the BH.

The connection with our ladder construction becomes particularly transparent in the vicinity of the potential maximum. At leading eikonal order, the barrier $V$ can be approximated by an inverted harmonic oscillator. Its exact scattering problem has the transmission probability 
\begin{equation} \label{GBF}
    \Gamma = \frac{1}{1+e^{-2\pi \mathcal{K}}} \, . 
\end{equation}
With our normalization, we obtain $\mathcal{K} \equiv
(\omega^2-V_0)/
{\sqrt{-2V_0''}} = \mathcal{E}^{\rm eik}/2$ for BHs, as the dimensionless energy measured from the top of the barrier. This immediately captures the expected limiting behavior: $\mathcal K=0$ corresponds to a wave at the top of the barrier and hence $\Gamma=1/2$, while $\mathcal K \gg 1$ and $\mathcal K \ll -1$ correspond to almost complete transmission and reflection, respectively. At the leading eikonal order, this corresponds also with the leading-order behavior for the BH reflectivity as computed in the WKB approximation~\cite{Iyer:1986np, Iyer:1986nq}. 

\begin{figure}[t]
\centering
\resizebox{\linewidth}{!}{%
\begin{tikzpicture}[
    >=Latex,
    font=\sffamily\large,
    titlebox/.style={
        rectangle,
        draw=black!80,
        fill=gray!10,
        thick,
        rounded corners=3pt,
        inner sep=6pt,
        align=center,
        text width=12.0cm
    },
    qnmbox/.style={
        rectangle,
        draw=blue!70!black,
        fill=blue!2,
        thick,
        rounded corners=3pt,
        inner sep=6pt,
        text width=3.55cm,
        align=left
    },
    reggebox/.style={
        rectangle,
        draw=red!70!black,
        fill=red!2,
        thick,
        rounded corners=3pt,
        inner sep=6pt,
        text width=3.55cm,
        align=left
    },
    gbfbox/.style={
        rectangle,
        draw=green!50!black,
        fill=green!2,
        thick,
        rounded corners=3pt,
        inner sep=6pt,
        text width=3.55cm,
        align=left
    }
]

\node[titlebox] (root) {
    \textbf{Ladder output}\\[2pt]
    \(\displaystyle
    \mathcal{E}(\omega,L)
    =
    2N+
    \sum_{k\geq1}
    \frac{\epsilon_{k}(N)}{L^k}
    \)
};


\node[reggebox, below=0.8cm of root.south] (regge) {
    \centering
    \textbf{\color{red!80!black}Regge poles}\\[-1pt]
    \color{red!80!black}
    \(\displaystyle N=n+\frac12\)
    \\[3pt]
    \hrule height 0.4pt\vspace{5pt}
    \raggedright

    \textbf{Fixed:} \(\omega, N\)\\[2pt]
    \textbf{Solve for:} \(L\)
};

\node[qnmbox, left=0.35cm of regge.west] (qnm) {
    \centering
    \textbf{\color{blue!80!black}QNMs}\\[-1pt]
    \color{blue!80!black}
    \(\displaystyle N=n+\frac12\)
    \\[3pt]
    \hrule height 0.4pt\vspace{5pt}
    \raggedright

    \textbf{Fixed:} \(L, N\)\\[2pt]
    \textbf{Solve for:} \(\omega\)

};

\node[gbfbox, right=0.35cm of regge.east] (gbf) {
    \centering
    \textbf{\color{green!50!black}GBFs}\\[4pt]
    \color{green!50!black}
    \(\displaystyle N=\mathcal{K}\)
    \\[12pt]
    \hrule height 0.4pt\vspace{5pt}
    \raggedright

    \textbf{Fixed:} \(\omega,L\)\\[2pt]
    \textbf{Solve for:} \(\mathcal{K}\)
};


\coordinate (junction) at ($(root.south)+(0,-0.30)$);

\draw[thick] (root.south) -- (junction);

\draw[->, thick, draw=blue!70!black]
    (junction) -| (qnm.north);

\draw[->, thick, draw=red!70!black]
    (junction) -- (regge.north);

\draw[->, thick, draw=green!50!black]
    (junction) -| (gbf.north);

\end{tikzpicture}%
}
\caption{
A unified description of the BH response from the ladder output. QNMs, Regge poles, and GBFs are obtained by different choices of the variables held fixed and
solved for.
}
\label{fig:unified_response}
\end{figure}
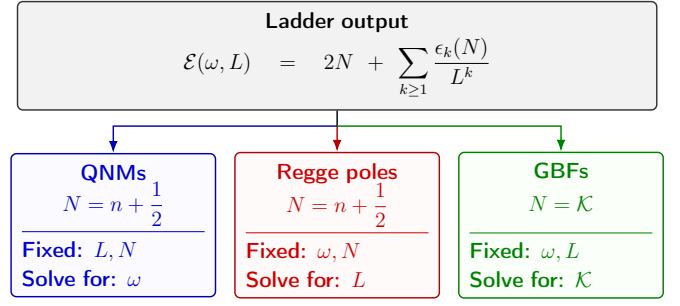

The crucial observation is that the inverted harmonic oscillator is only the leading term in our expansion. Once anharmonic corrections to the potential are included, the functional form of Eq.~\eqref{GBF} remains unchanged; rather, the anharmonicity modifies the effective barrier-top parameter $\mathcal{K}$. By comparing with the leading order, notice that $\mathcal{K}$ coincides with $N$, if we interpret $N$ as a function of $\{\omega, L\}$. Indeed, in order to determine the BH GBFs, we can solve the same equation as we did for the QNMs and the Regge poles, but analytically extending $N \to \mathcal{K}$ and keeping $\{\omega, L\}$ as free variables: 
\begin{equation}
    \mathcal{E}(\omega,L)= 2\, \mathcal{K} + \sum_k \frac{\epsilon_{k}(\mathcal{K})}{L^k} \, .
\end{equation}
This produces a very accurate approximation for the GBF (and for the BH reflectivity). For reference, we write here the first few terms:
\begin{equation} \label{GBFK}
    \begin{aligned}
        \mathcal{K} =&\, \frac{\mathrm{1}}{L}\left[\frac{27 \tilde{\omega}}{2} - \frac{\beta}{3} + \frac{115}{432}\right]  \\
        &+ \frac{1}{L^3}\Bigl[-\frac{1863}{16}\,\tilde{\omega}^2 + \left(\frac{265}{1152} - \frac{\beta}{4}\right) \tilde{\omega} \\
        &+ \frac{19\,\beta^2}{324} - \frac{4177\,\beta}{46656} + \frac{1356539}{40310784}\Bigr] + \mathscr{O}(L^{-5}) \, , 
    \end{aligned}
\end{equation}
where $\tilde{\omega} \equiv (\varpi^2-L^2)/27$. Notice that at alternating orders in $L$, the high-frequency behavior of $\mathcal{K}$ changes. We will later see how this alternating behavior can be softened by considering resummations of the post-eikonal expansion for $\mathcal{K}$, which lead to a significantly more accurate approximation. We summarize our ladder-based approach for finding QNMs, Regge poles, and GBFs in Fig.~\ref{fig:unified_response}. 

\section{Results and Accuracy}\label{sec:results}

Using our recursive ladder implementation, we have extended the asymptotic QNM series in Eq.~\eqref{QNMser} up to $\mathscr{O}(L^{-18})$, and computed the corresponding series for Regge poles and GBFs as well following the prescription outlined above. Since the resulting expressions are too lengthy to present here, we make them available in the accompanying ancillary files.

We now study the accuracy and convergence properties of these asymptotic expansions. It is well known that the WKB expansion for QNMs shows improved convergence under resummation of high WKB orders~\cite{Konoplya:2019hlu, Hu:2026cyq}. Motivated by this, we also investigate the extent to which the performance of the asymptotic expansions can be improved through the following resummation techniques:

\noindent
{\bf (i) Pad\'e resummation}: The procedure is most conveniently described for a generic quantity $F(L)$ admitting a large-$L$ expansion of the form
\begin{equation} \label{Padeser}
    F(L) = \sum_{k=k_0}^{\infty} c_k\, L^{-k},
\end{equation}
where $k_0$ can be negative and the coefficients $c_k$ may, in general, be complex. In the applications considered here, we are interested in the cases with $k_0=-1$. Introducing $g \equiv 1/L$, the corresponding asymptotic expansion can be written as 
\begin{equation*}
    F(g) = \frac{1}{g} \left(c_{-1}+c_0\, g + c_1\, g^2+\cdots\right).
\end{equation*}
We first truncate the series at a finite order $T$ to obtain $F^{(T)}(g)=R^{(T+1)}(g)/g$, where $R^{(T+1)}(g)=\sum_{j=0}^{T+1} c_{j-1}\, g^j$ is regular at $g=0$. Rather than constructing a rational approximation directly for the singular function $F(g)$, we construct the Pad\'e approximant for its regular series $R^{(T+1)}(g)$.

For integers $\{m,n\}$ such that $m+n=T+1$, we define the $[m,n]$-Pad\'e approximant to $R^{(T+1)}(g)$ as
\begin{equation} \label{Pade1}
    \mathcal{R}^{m}_{n}(g) = \frac{\sum_{i=0}^{m} p_i\, g^i}{\sum_{j=0}^{n} q_j\, g^j},\quad q_0=1,
\end{equation}
with the coefficients $\{p_i,q_j\}$ determined by requiring matching with the Taylor expansion up to $g^{T+1}$. The resulting Pad\'e-approximant to the original series $F(g)$ is therefore
\begin{equation} \label{Pade2}
    \mathcal{F}^{m}_{n}(L) = L\, \mathcal{R}^{m}_{n}(1/L).
\end{equation}
Since a Pad\'e approximant depends both on the number of available asymptotic coefficients and on how they are distributed between the numerator and denominator, we do not restrict our analysis to a single choice of approximant. Instead, we will vary the truncation order $T$ and the numerator order $m$, allowing us to investigate the sensitivity of the resummed results to the choice of Pad\'e approximant.

\noindent
{\bf (ii) Borel-Pad\'e resummation}: We also employ the Borel-Pad\'e resummation technique, where we notice that a direct Borel transformation of $F(g)$ is not useful because of the explicit $1/g$ term. We therefore first separate the leading eikonal contribution
\begin{equation} \label{BP1}
    \widetilde{F}^{T}(g) \equiv F(g)-\frac{c_{-1}}{g}-c_0 = \sum_{k=1}^{T} c_k\, g^k
\end{equation}
and apply the Borel transformation only on the subleading part
\begin{equation} \label{BP2}
    \mathcal{B}[\widetilde F](s)= \sum_{k=1}^{T} \frac{c_k}{k!}\, s^k\, .
\end{equation}
Then, we implement an $[m,T-m]$-Pad\'e approximation to this Borel-transformed quantity. Finally, the resummed quantity is obtained through the inverse Borel-Laplace transform as
\begin{equation} \label{BP3}
    F^{m}_{T-m,\, BP}(L) = c_{-1}\, L + c_0 + \int_0^\infty dt\, e^{-t}\, \widetilde{\mathcal{F}}^{m}_{T-m}(t/L). 
\end{equation}
As with the ordinary Pad\'e resummation, we will vary the truncation order ($T$) and the Pad\'e order ($m$) to assess the accuracy of the resulting resummed quantities.

\subsection{QNM Frequencies}

\begin{figure}[t]
    \centering
    \includegraphics[width=\columnwidth]{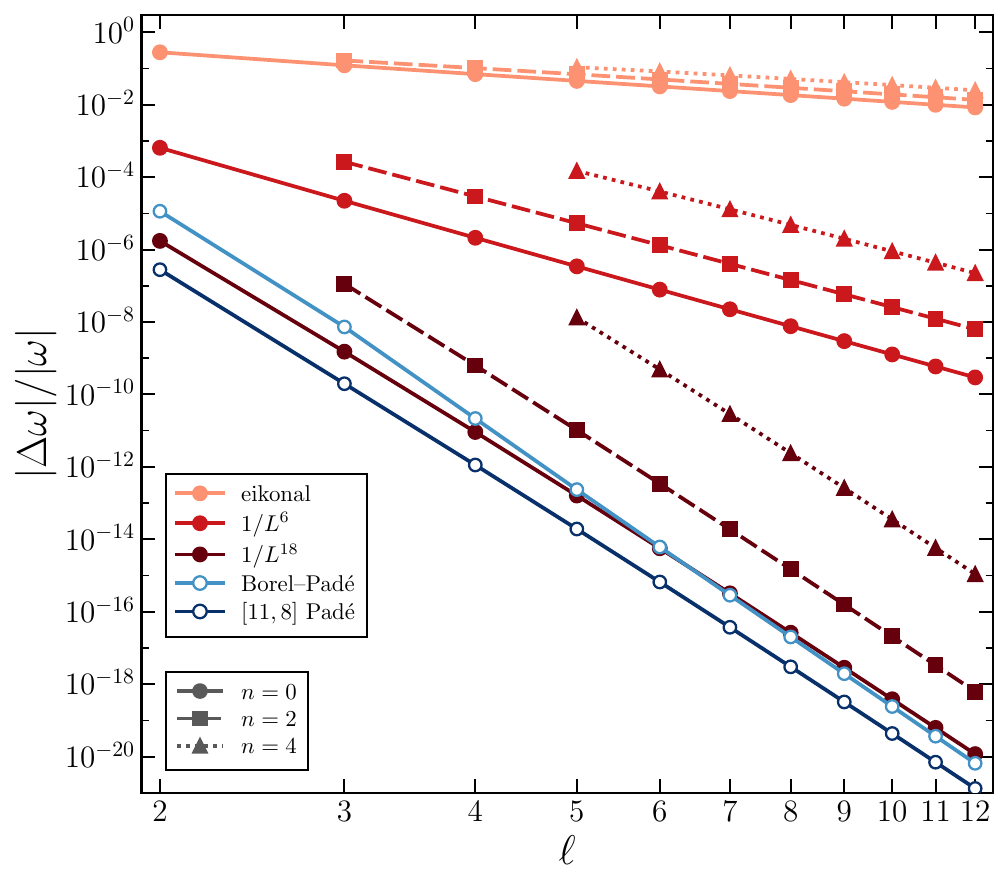}
    \caption{Residual in the gravitational QNM frequency between the exact values computed from Leaver's method and the sub-eikonal approximation truncated to $\mathscr{O}(L^0)$, $\mathscr{O}(L^{-6})$, and $\mathscr{O}(L^{-18})$. Different markers represent different overtone numbers $n=\{0,\,2,\,4\}$. We do not show the overtone numbers with $n>\ell$, as we do not expect the approximation to be very accurate in that regime. Notice that including up to $\mathscr{O}(L^{-18})$ corrections leads to a prediction for the fundamental $\ell=2$ mode that is accurate to within $10^{-6}$. In addition, we show in blue two different kinds of resummation at $\mathscr{O}(L^{-18})$: Borel-Pad\'e (light blue), and a $[11,\,8]$-Pad\'e approximant (dark blue). The latter shows a slight improvement compared to the bare post-eikonal expansion.}
    \label{fig:accuracy}
\end{figure}

We first assess the accuracy and convergence properties of the $1/L$-expansion for the QNM spectrum. As we include more terms in the series, i.e., increase the truncation order $T$, the asymptotic series converges closer to the QNMs obtained via Leaver's method~\cite{Leaver:1985ax}. The improvement in accuracy is more significant for higher $L$ values, in accordance with the approximation scheme. Remarkably, by going to sufficiently high order, we can calculate very accurately the frequency of low-$L$ QNMs. This proves that this expansion method is also effective beyond the eikonal regime, especially for the fundamental modes ($n=0$). For instance, the $\mathscr{O}(L^{-18})$ approximation derived here for the fundamental gravitational QNM is accurate to within $10^{-6}$, i.e., it is indistinguishable from the exact value for observational purposes. 

\begin{figure}[t]
    \centering
    \includegraphics[width=\columnwidth]{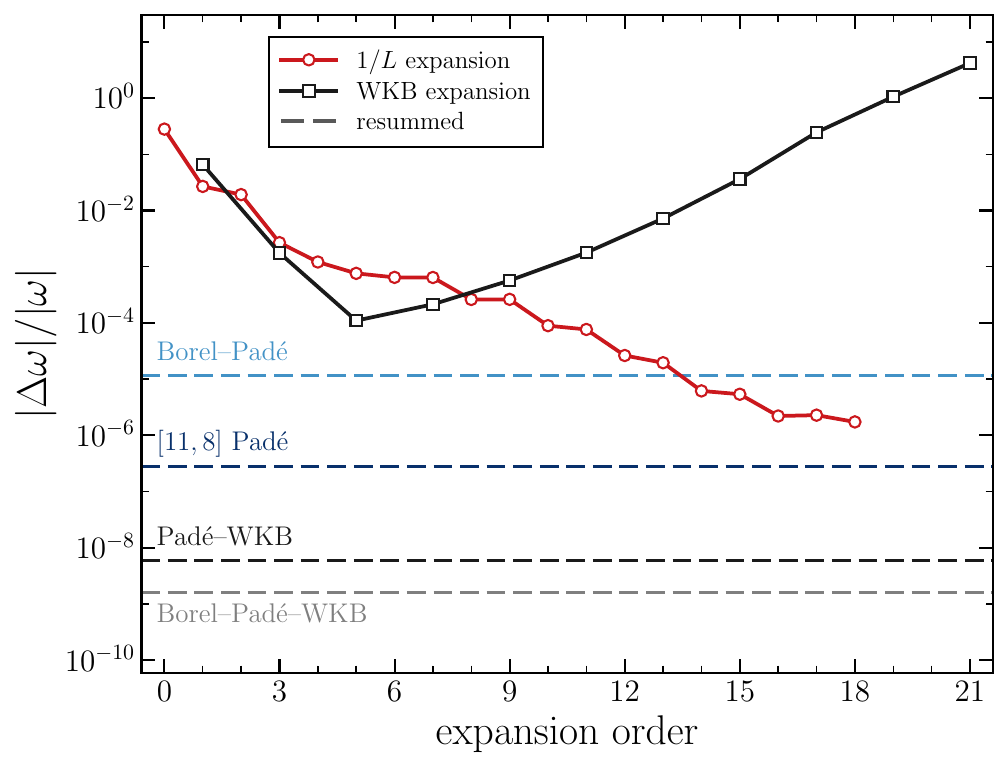}
    \caption{Accuracy of the eikonal expansion compared with the WKB expansion of Ref.~\cite{Hu:2026cyq} for the fundamental gravitational mode, as a function of the expansion order. While the bare WKB expansion does not converge beyond fourth order, the bare eikonal expansion improves as one includes higher orders. However, the resummed WKB method seems to outperform the resummed post-eikonal series by a few orders of magnitude.}
    \label{fig:wkb_vs_eikonal}
\end{figure}

\begin{figure}[t]
    \centering
    \includegraphics[width=\columnwidth]{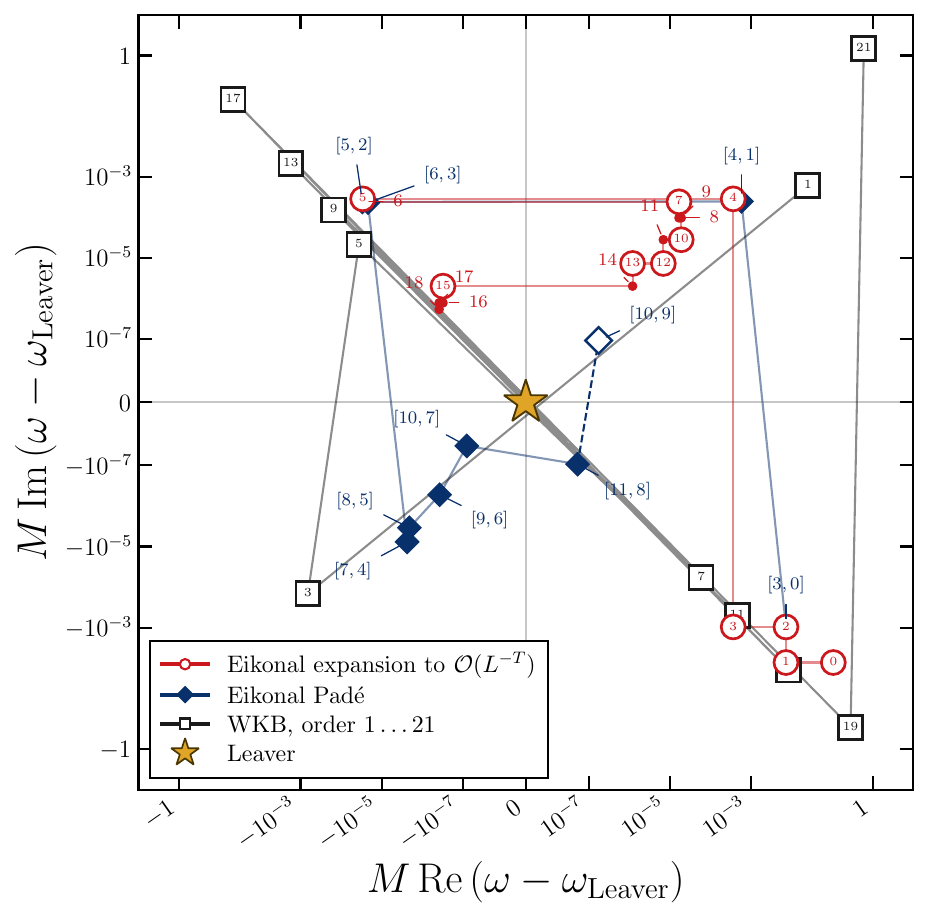}
    \caption{Complex-plane trajectories of the fundamental gravitational mode frequency, obtained from the post-eikonal expansion (red circles), its Pad\'e resummation (blue diamonds), and the WKB approximation of~\cite{Hu:2026cyq} (black squares). We observe a step-like convergence of the post-eikonal approximation to the exact value, as opposed to the oscillating (and divergent) behavior of the WKB series. The star at the center marks the QNM frequency from Leaver's algorithm.}
    \label{fig:trajectory}
\end{figure}

As the overtone number grows, the coefficients in the QNM series expansion~\eqref{QNMser} grow as $N\, (N/L)^k$. This means that the series will converge at a slower rate to the exact value for higher overtones. In particular, we should not expect the series to be very accurate for high overtones with overtone number $n \geq \ell$. 

In Fig.~\ref{fig:accuracy} we summarize the convergence of the post-eikonal QNM expansion. We consider only gravitational QNMs (spin $s=-2$), and overtone numbers $n=\{0,\,2,\,4\}$. In order to have a meaningful comparison at high-$\ell$ values, we implemented an arbitrary precision Leaver's method to determine the QNMs with very high accuracy. This is what allows us to confidently report relative errors of the order of $10^{-20}$ for our eikonal approximation with $\ell=12$ at $\mathscr{O}(L^{-18})$. The improvement achieved by going to higher orders in the post-eikonal expansion is remarkable. We have also studied two different resummation strategies. The most accurate results are obtained with a $[10,\,8]$-Pad\'e approximant (dark blue line), whereas the Borel-Pad\'e resummation scheme performs worse than the asymptotic expansion itself. Even the Pad\'e resummation does not lead to a dramatic improvement. We discuss this in more detail below. 

It is quite illustrative to compare the eikonal expansion carried out in this work with the traditional WKB expansion~\cite{Iyer:1986np, Iyer:1986nq}, and its resummations~\cite{Hu:2026cyq}. Our findings are summarized in Fig.~\ref{fig:wkb_vs_eikonal}. First, we note that the bare WKB expansion does \emph{not} converge: going to higher WKB orders does not improve the accuracy of the approximation. In fact, the optimal truncation order seems to be $T_{\rm WKB}^{\rm optimal}=4$. On the other hand, the eikonal expansion performed here converges uniformly. The main difference lies in the effectiveness of resummation strategies. As mentioned before, the Pad\'e or Borel-Pad\'e resummations do not significantly improve the accuracy of our eikonal approximation. On the other hand, the resummation of the WKB series leads to a dramatic improvement, achieving results that are two orders of magnitude more accurate than the eikonal expansion. We emphasize, however, that the difference between these two approximations is way far beyond what any current, planned, and hypothetical GW detector will be able to resolve in the foreseeable future.  

We can ask why resummation is so useful for the WKB expansion, while it does not seem to help the eikonal approximation scheme. The reason can be found in Fig.~\ref{fig:trajectory}. Consider the WKB expansion (black squares). Being a divergent series, the successive orders oscillate away from the exact QNM frequency. Resummation, in this case, ``averages'' away these oscillations, and finds a very accurate result. On the other hand the red trajectory followed by the eikonal expansion resembles a step-like structure, steadily approaching the QNM frequency. We also show in blue markers the different Pad\'e approximants, for which we find a slightly better performance with an off-diagonal Pad\'e $\mathcal{F}^{11}_8$, as opposed to the (near)-diagonal $\mathcal{F}^{10}_9$.

\subsection{Regge Poles}

\begin{figure}[t]
    \centering
    \includegraphics[width=\columnwidth]{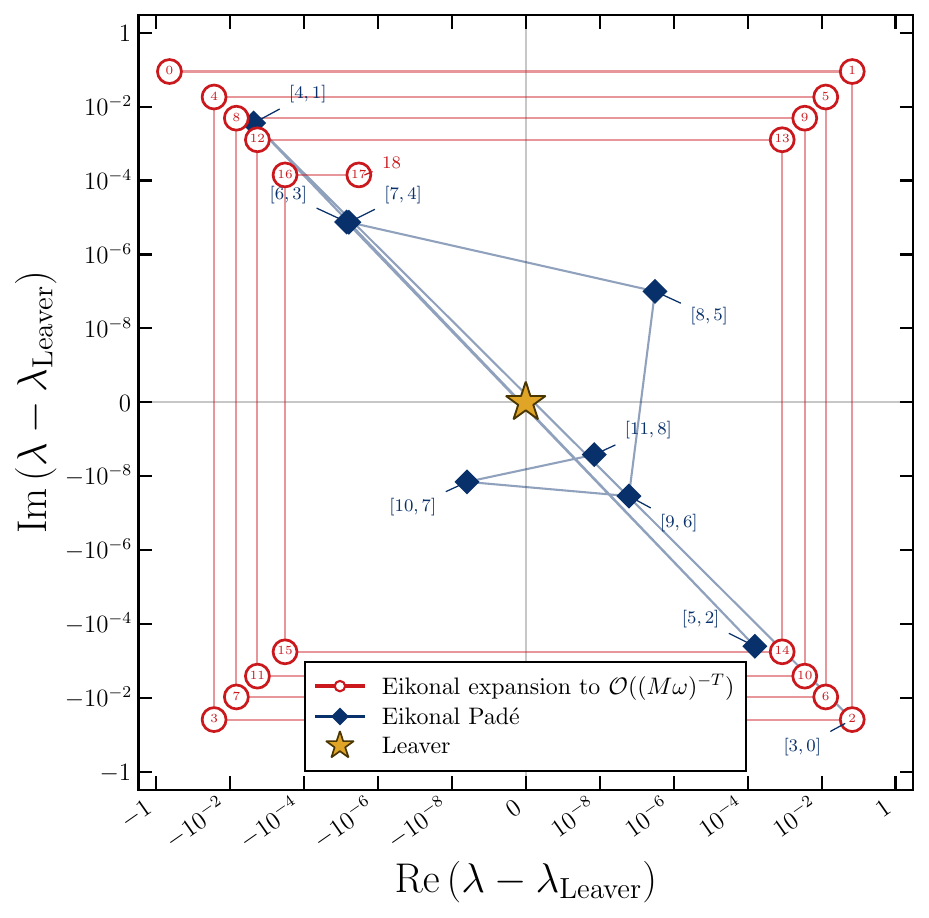}
    \caption{Complex-plane Trajectories of the fundamental Regge pole $\lambda$ for $M\omega =0.5$ and $n=0$. The red line corresponds to truncating the eikonal expansion at order $T \in \{0,\dots, 18\}$, and the blue line to the Pad\'e-resummed series. The star at the center marks the value of the Regge pole found through Leaver's algorithm.}
    \label{fig:regge}
\end{figure}

Next, we study the accuracy and convergence properties for the Regge poles $\lambda_\omega^{(n)}$, and compare the performance of the unresummed asymptotic expansion with its Pad\'e-resummed counterpart. As an illustration, in Fig.~\ref{fig:regge} we show the trajectories traced by the eikonal approximation and its Pad\'e resummation in the complex $\lambda_\omega$-plane for the fundamental gravitational Regge pole at $M\omega=0.5$.
The convergence behavior is qualitatively different from that observed for the QNM spectrum. In the unresummed case, the successive approximations approach the exact result through a characteristic sequence of overshoots and undershoots in the real and imaginary parts, producing a spiral-like trajectory toward the target. By contrast, the Pad\'e-resummed series follows a different trajectory as the resummation order is increased, reflecting the distinct manner in which the rational approximant reorganizes the available asymptotic information. Unlike the QNM expansion, the alternating behavior of the series means that resummation is effective and leads to a much more accurate result for the Regge poles. 

Because the asymptotic expansion contains inverse powers of $M\omega$, it becomes singular in the limit $M\omega \to 0$, and its accuracy correspondingly deteriorates in the immediate vicinity of this point. Nevertheless, the expansion remains remarkably accurate at frequencies away from this singular limit. In particular, both the unresummed and Pad\'e-resummed series continue to provide good approximations down to frequencies as low as $M\omega = 0.5$, demonstrating that the large-$M\omega$ expansion remains useful over a substantially broader frequency range than might be inferred from its formal singularity at $M\omega=0$.

\subsection{Greybody Factors}

\begin{figure}[t]
    \centering
    \includegraphics[width=\columnwidth]{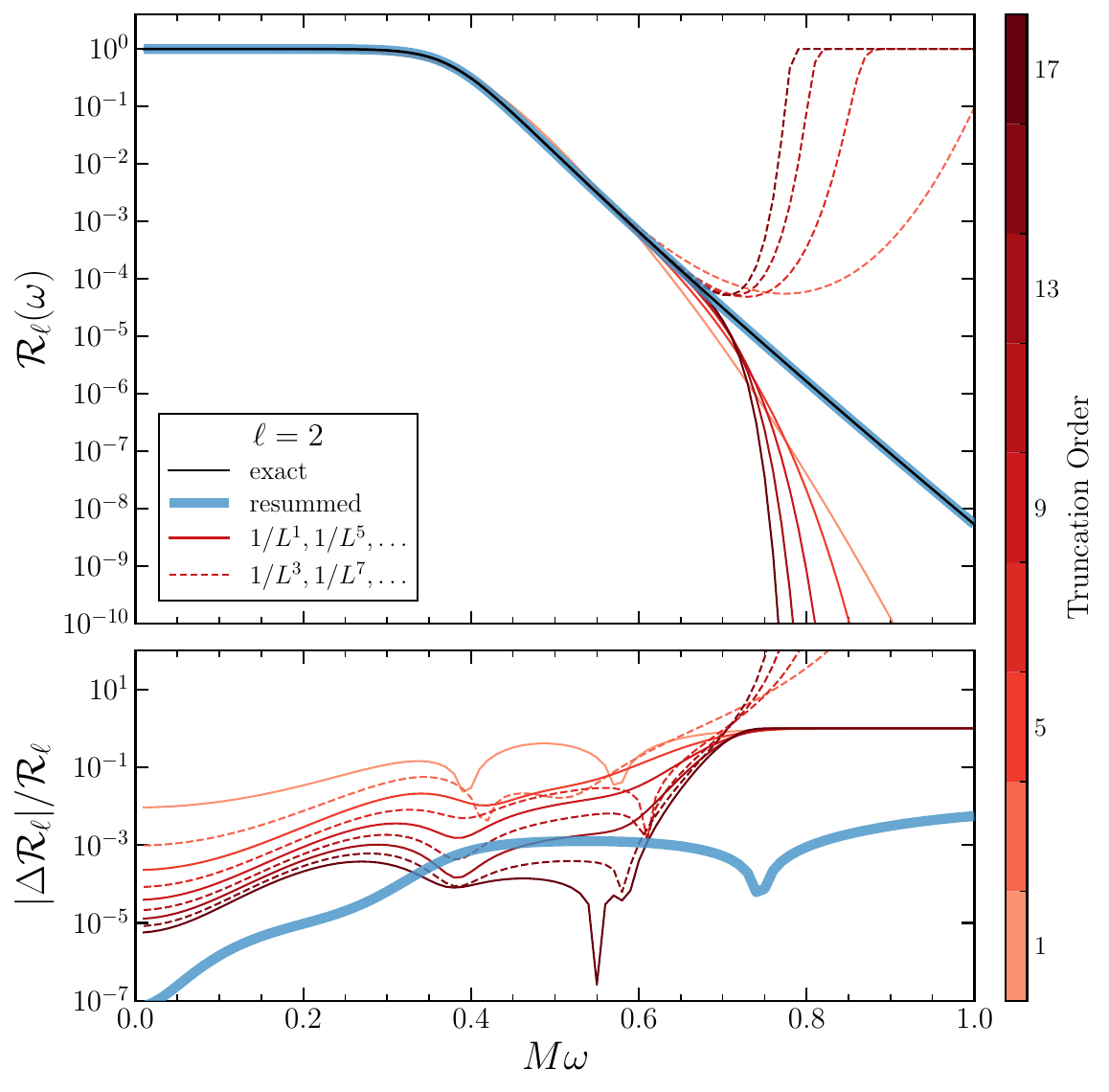}
    \caption{BH reflectivity for the gravitational quadrupole $\ell=2$ mode. The exact values (black) agree to sub-percent accuracy with the resummed eikonal expansion (blue). The different red lines show the reflectivity obtained from the truncated series, where the truncation order is indicated by the color bar, with darker hues indicating higher truncation order.}
    \label{fig:gbf_2}
\end{figure}

\begin{figure}[t]
    \centering
    \includegraphics[width=\columnwidth]{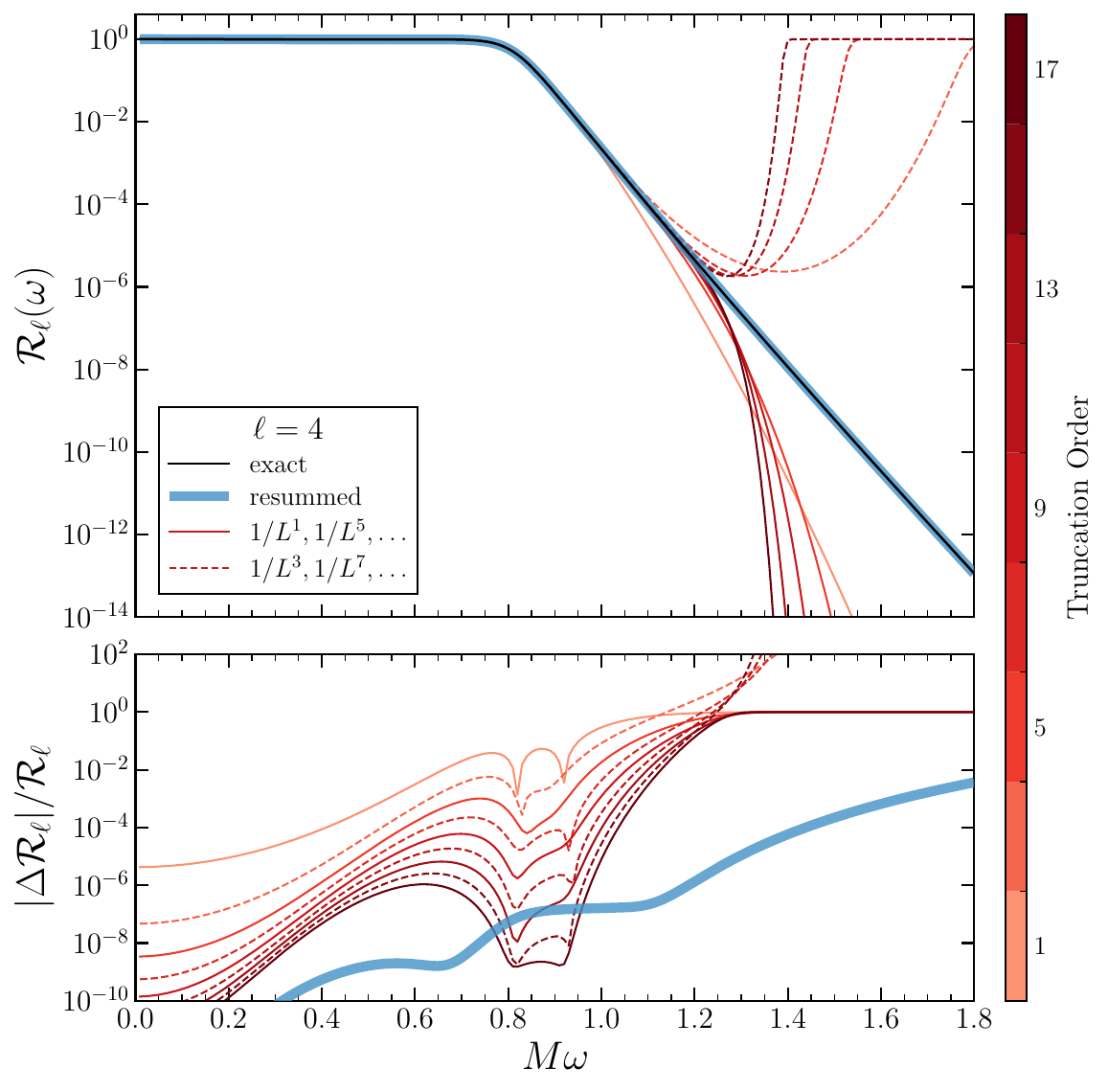}
    \caption{Same as in Fig.~\ref{fig:gbf_2} but for the $\ell=4$ mode. As expected, the agreement between the eikonal approximation and the exact value is much better in this case.}
    \label{fig:gbf_4}
\end{figure}

Finally we study the accuracy of the eikonal expansion applied to the GBFs. The ``exact'' values for the BH reflectivity are obtained using the Teukolsky package of the \texttt{Black Hole Perturbation Toolkit}~\cite{BHPToolkit}. Recall that the GBF is obtained from the BH reflectivity as $\Gamma_\ell = 1-\mathcal{R}_\ell$. We use the BH reflectivity in the figures because it shows more clearly the convergence at high frequencies, and because this is the quantity imprinted in the GW waveform of binary BHs~\cite{Oshita:2023cjz, Okabayashi:2024qbz, Rosato:2024arw, Rosato:2025ulx, Rosato:2026apq}. 

The expansion of $\mathcal{K}(\omega, L)$ in Eq.~\eqref{GBFK} can be written formally as 
\begin{equation}
    \mathcal{K} = \sum_{k=1}^T \frac{P_k(\tilde{\omega})}{L^{2k-1}} \, , 
\end{equation}
where each $P_k$ is a polynomial in $\tilde{\omega}$. This is such that $\tilde{\omega}=0$ corresponds to incident waves whose energy matches exactly the peak of the potential barrier (in the eikonal limit). For $\tilde{\omega}>0$, the series expansion does not converge (see Figs.~\ref{fig:gbf_2} and \ref{fig:gbf_4} below). We consider a resummation which replaces each polynomial $P_k \mapsto \tilde{P}_k$ by its diagonal (or near-diagonal) Pad\'e approximant in $\tilde{\omega}/L^2$, making use of all the available asymptotic information. The resummed series is then 
\begin{equation}
    \mathcal{K}_{\rm resum} = \sum_{k=1}^T \frac{\tilde{P}_k(\tilde{\omega}/L^2)}{L^{2k-1}} \, .
\end{equation}
We show the accuracy of the series truncated to order $T$ and its resummation (including up to $17$th order) in Fig.~\ref{fig:gbf_2} (for $\ell=2$) and Fig.~\ref{fig:gbf_4} (for $\ell=4$). As expected, for $ \varpi > L$, the series stops converging accurately. In fact, we find a similar behavior as for the WKB series applied to the QNMs, or for the Regge poles --- the truncated value of $\mathcal{K}$ alternatively overshoots (solid lines) or undershoots (dashed lines), leading to a reflectivity that goes to $1$ or $0$ rapidly. This alternating behavior is remedied by the resummation, which leads to a very accurate approximation of the BH reflectivity, producing sub-percent level accuracy even at high frequencies. The BH reflectivity decays exponentially at high frequencies, so one must read the corresponding relative error residuals with a certain caution. 

We also highlight that the eikonal approximation itself \emph{does} converge, as expected, at low frequencies. At $17$th order we achieve an accuracy better than $10^{-3}$ for all frequencies below the fundamental QNM frequency for $\ell=2$, and better than $10^{-6}$ for $\ell=4$.

\section{Discussion}\label{sec:conclusions}
It has long been recognized that the near-horizon ladder symmetry provides a geometric perspective on the tidal response of BHs and, in particular, on the vanishing of their static Love numbers in general relativity~\cite{Hui:2021vcv, Charalambous:2021kcz, BenAchour:2022uqo, Charalambous:2022rre, Sharma:2024hlz, Ghosh:2026vig}. In contrast, this work exploits the near-LR ladder symmetry of the same perturbation equations in the eikonal regime, and demonstrates that it likewise governs a broad class of BH scattering observables, including QNMs, Regge poles, and GBFs. By appropriately reorganizing the perturbation equations, we map the problem onto a harmonic oscillator supplemented by anharmonic corrections organized as a series in inverse angular momentum ($L^{-1}$). This mapping allows us to apply standard perturbation theory and recursively determine the sub-eikonal QNM frequencies in terms of the eikonal ladder operators. We further extend this framework to compute Regge poles and GBFs.

The resulting asymptotic QNM series provides an accurate description of the Schwarzschild spectrum. In contrast to the corresponding WKB expansion, our series exhibits uniform convergence with increasing truncation order $\mathscr{O}(L^{-T})$, while Pad\'e resummation procedures yield only modest further improvements for the QNMs. This demonstrates that the ladder-based expansion remains effective beyond the strict eikonal regime, including for the fundamental modes with $n=0$. For example, the $\mathscr{O}(L^{-18})$ approximation for the fundamental gravitational QNM reproduces the exact result to seven decimal places, making the difference negligible at the level relevant for GW observations. For Regge poles and GBFs, particularly in the high-frequency regime, resummation becomes more important: the asymptotic series can achieve substantially improved accuracy when combined with Pad\'e approximants.

There are several natural directions in which our framework can be extended. Perhaps the most immediate is its application to rotating Kerr BHs. We expect such an extension to be feasible, at least for equatorial QNMs, although a central challenge is the simultaneous treatment of the radial and angular Teukolsky equations and their associated separation constant. Another natural avenue is to apply the expansion scheme to BHs in theories beyond general relativity, for instance within the parametrized-potential framework of Ref.~\cite{Tang:2025qaq} (see also Refs.~\cite{Cardoso:2019mqo,McManus:2019ulj,Li:2022pcy,Volkel:2022aca,Volkel:2022khh,Hirano:2024fgp,Cano:2024jkd}). The universal near-LR structure, characterized by a leading harmonic term supplemented by systematically organized subleading corrections, provides strong motivation for the applicability of our approach to such systems.

A more fundamental open question concerns the connection between our perturbative framework and genuinely non-perturbative calculations of QNMs. Rather than treating the anharmonic corrections order-by-order in $L^{-1}$, one could attempt to resum the resulting series and solve the corresponding anharmonic oscillator problem directly. In this setting, the underlying algebra is no longer simply $\mathfrak{sl}(2,\mathbb{R})$, but rather a deformation of this algebra. It is conceivable that, upon incorporating the full tower of corrections, the resulting quantum-mechanical problem could admit a description in terms of Seiberg-Witten curves~\cite{Aminov:2020yma}. Exploring this possible connection would provide an intriguing bridge between BH perturbation theory and non-perturbative quantum-mechanical methods, although a detailed investigation is beyond the scope of the present work.

\section*{Acknowledgements}

We thank Jierui Hu for sharing the data of Ref.~\cite{Hu:2026cyq}. 
R.G., D.P. J.R.-Y. and E.B. are supported by NSF Grants No.~AST-2606672, No.~PHY-2513337, No.~PHY-090003, and No.~PHY-20043, by John Templeton Foundation Grant No.~62840, by the Simons Foundation [MPS-SIP-00001698, E.B.], by the Simons Foundation International [SFI-MPS-BH-00012593-02], and by Italian Ministry of Foreign Affairs and International Cooperation Grant No.~PGR01167.
R.G. is supported by the Fulbright Nehru Postdoctoral Research Fellowship (Award No.3174/FNPDR/2025) from the United States-India Educational Foundation.
This work makes use of the \texttt{Black Hole Perturbation Toolkit}~\cite{BHPToolkit}.
Part of this work was carried out at the Advanced Research Computing at Hopkins (ARCH) core facility (\url{https://www.arch.jhu.edu/}), which is supported by the NSF Grant No. OAC-1920103.
The Tycho supercomputer hosted at the SCIENCE HPC center at the University of Copenhagen was used for supporting this work.

\bibliography{biblio}

\end{document}